\documentclass[10pt,letterpaper]{article}
\pdfoutput=1

\usepackage{jcappub}
\usepackage{amsmath,amssymb,color}
\usepackage{enumerate,xstring}
\usepackage{jcappub}
\usepackage{amsmath}
\usepackage{amsfonts}
\usepackage{amssymb}
\usepackage{graphicx}
\usepackage{epsfig}
\usepackage{color}
\usepackage{multirow}
\usepackage{graphicx}
\usepackage{bigints}
\usepackage{todonotes,bm,etoolbox}
\usepackage{calligra}
\usepackage{hyperref}
\usepackage{tabularx}
\usepackage{booktabs}
\usepackage{subfigure}

\usepackage{tikz}

\def\ba#1\ea{\begin{align}#1\end{align}}
\def\bea{\begin{eqnarray}}
\def\eea{\end{eqnarray}}
\def\be{\begin{equation}}
\def\ee{\end{equation}}

\def\({\left(}
\def\){\right)}
\def\[{\left[}
\def\]{\right]}

\def\<{\left\langle}
\def\>{\right\rangle}

\def\comment#1{}

\def\eps{\epsilon}

\renewcommand{\v}[1]{\bm{#1}}

\def\vx{\v{x}}
\def\vk{\v{k}}

\def\vtheta{\v{\theta}}

\def\vM{\v{M}}
\def\vD{\v{D}}

\newcommand{\perm}[1]{ \expandafter\ifstrempty\expandafter{#1} {\mbox{perm.}} {\mbox{$#1$ perm.}} }

\def\O{\mathcal{O}}
\def\M{\mathcal{M}}

\def\R{\mathcal{R}}

\newcommand{\fnl}{f_\textnormal{\textsc{nl}}}

\newcommand{\bphi}{b_\phi}
\newcommand{\bsigma}{b_\sigma}

\newcommand{\A}{\mathcal{A}}

\newcommand{\fidu}{{\rm Fiducial}}
\newcommand{\highcip}{{\rm High}\sigma}
\newcommand{\lowcip}{{\rm Low}\sigma}

\definecolor{RedWine}{rgb}{0.743,0,0}
\definecolor{RoyalBlue}{rgb}{0.25,.41,.88}
\definecolor{ForestGreen}{rgb}{.13,.54,.13}
\definecolor{Goldenrod}{rgb}{.85,.65,.13}

\newcommand{\bq}{\begin{eqnarray}}
\newcommand{\eq}{\end{eqnarray}}

\def\pathtofigs{./}

\title{\huge Constraints on compensated isocurvature perturbations from BOSS DR12 galaxy data}

\author[a,b]{Alexandre Barreira}

\affiliation[a]{\small Excellence Cluster ORIGINS, Boltzmannstra\ss e 2, 85748 Garching, Germany}
\affiliation[b]{\small Ludwig-Maximilians-Universit\"at, Schellingstra\ss e 4, 80799 M\"unchen, Germany}

\emailAdd{alex.barreira@origins-cluster.de}

\date{\today}

\abstract{We use the BOSS DR12 galaxy power spectrum to constrain compensated isocurvature perturbations (CIP), which are opposite-sign primordial baryon and dark matter perturbations that leave the total matter density unchanged. Long-wavelength CIP $\sigma(\vx)$ enter the galaxy density contrast as $\delta_g(\vx) \supset \bsigma\sigma(\vx)$, with $\bsigma$ the linear CIP galaxy bias parameter. We parameterize the CIP spectra as $P_{\sigma\sigma} = A^2P_{\R\R}$ and $P_{\sigma\R} = \xi\sqrt{P_{\sigma\sigma}P_{\R\R}}$, where $A$ is the CIP amplitude and $\xi$ is the correlation with the curvature perturbations $\R$. We find a significance of detection of $A\bsigma \neq 0$ of $1.8\sigma$ for correlated ($\xi = 1$) and $3.7\sigma$ for uncorrelated ($\xi = 0$) CIP. Large-scale data systematics have a bigger impact for uncorrelated CIP, which may explain the large significance of detection. The constraints on $A$ depend on the assumed priors for the $\bsigma$ parameter, which we estimate using separate universe simulations. Assuming $\bsigma$ values representative of all halos we find $\sigma_A = 145$ for correlated CIP and $\sigma_{|A|} = 475$ for uncorrelated CIP. Our strongest uncorrelated CIP constraint is for $\bsigma$ representative of the $33\%$ most concentrated halos, $\sigma_{|A|} = 197$, which is better than the  current CMB bounds $|A| \lesssim 360$. We also discuss the impact of the local primordial non-Gaussianity parameter $\fnl$ in CIP constraints. Our results demonstrate the power of galaxy data to place tight constraints on CIP, and motivate works to understand better the impact of data systematics, as well as to determine theory priors for $\bsigma$.}

\begin{document}

\maketitle

\section{Introduction}
\label{sec:intro}

Understanding the physics of {\it primordial inflation} is one of the main goals of modern cosmology. The simplest models postulate the existence of a single scalar particle species that drove inflation and whose quantum fluctuations planted the primordial density perturbations that seeded cosmic structure formation. A key prediction of these models is that the perturbations are {\it adiabatic}, meaning the density fluctuations of all particle species are proportional to the same curvature perturbation field $\R({\vx})$. The departure from adiabaticity is dubbed {\it isocurvature}, and its detection would carry profound consequences to fundamental physics: it would rule out single-field inflation and indicate the early Universe featured multi-field dynamics. Traditionally, isocurvature in species $i$ is defined w.r.t.~the photon ($\gamma$) number density as $S_{i\gamma} = \delta n_i/\bar{n}_i - \delta n_\gamma/\bar{n}_\gamma$, where $\bar{n}_i$ is the mean number density and $\delta n_{i}$ its fluctuation. Most forms of isocurvature are very tightly constrained by the cosmic microwave background (CMB) data: the Planck satellite set an upper limit of $2\%$ ($2\sigma$) for the contribution of several isocurvature modes to the CMB temperature power spectrum \cite{2020AA...641A..10P}.

\vspace{1mm}
Here, we focus on a form of isocurvature known as {\it baryon-cold dark matter compensated isocurvature perturbations} (CIP) \cite{2003PhRvD..67l3513G, grin/dore/kamionkowski, 2014PhRvD..89b3006G}, which is interesting in that it is remarkably poorly constrained by the CMB data. The CIP field is characterized by perturbations in the baryons ($b$) that are compensated by opposite sign perturbations in the cold dark matter (CDM) ($c$) that yield zero total matter ($m$) isocurvature. Concretely, it is a combination of baryon and CDM isocurvature modes described as 
\bq\label{eq:cipdef}
S_{c\gamma} = - (\Omega_b/\Omega_c)S_{b\gamma} \Longrightarrow S_{m\gamma} = (\Omega_c/\Omega_m)S_{c\gamma} + (\Omega_b/\Omega_m)S_{b\gamma} = 0,
\eq
where $\Omega_i$ is the cosmic fractional energy of species $i$.\footnote{We assume throughout a regime where both baryons and CDM are already non-relativistic, and exclude also neutrinos from what we call {\it total matter}.} Thus, CIP do not alter gravitational potentials to linear order, and impact the CMB only through a spatial modulation of the plasma sound speed, which is a second-order effect. The latest CMB constraints still allow the power spectrum of CIP to be $5$ orders of magnitude larger (!) than the power spectrum of the adiabatic perturbations $\R(\vx)$ \cite{2016PhRvD..93d3008M, 2017PhRvD..96h3508S, 2017JCAP...04..014V, 2020AA...641A..10P, 2021PhRvD.104j3509L}. The CIP can be produced in multi-field inflation models like the {\it curvaton} \cite{2015PhRvD..92f3018H}, as well as in certain baryogenesis scenarios \cite{2016JCAP...08..052D}.

\vspace{1mm}
The late-time distribution of galaxies can also be used to probe CIP. For example, long-wavelength CIP impact the formation of galaxies through two main effects \cite{2020JCAP...02..005B}. First, they modify the local baryon fraction, which naturally impacts the astrophysics of gas accretion, star formation and stellar/black hole feedback. Second, the extra baryons cause also the total matter power spectrum to grow less on $k > k_{\rm eq}\approx 0.02 h/{\rm Mpc}$ in between matter-radiation equality and recombination. This second effect affects not only the formation of galaxies, but of all structures including halos. Technically, this means that bias expansions of the galaxy density contrast $\delta_g(\vx)$ \cite{biasreview} must include contributions such as $\delta_g(\vx) \supset \bsigma\sigma(\vx)$, where $\sigma$ denotes a CIP and $\bsigma$ is a bias parameter that quantifies how many more galaxies form inside a large-scale CIP. This bias parameter was studied using numerical simulations in Refs.~\cite{2020JCAP...02..005B, Voivodic:2020bec, 2021JCAP...03..023K}, and the idea to constrain CIP using this contribution to the bias expansion has been explored with forecasts in Refs.~\cite{2019PhRvD.100j3528H, 2020JCAP...07..049B, 2021PhRvD.103h3519S, 2022arXiv220802829K}. The phenomenology of these constraints is similar to the constraints on the local primordial non-Gaussianity (PNG) parameter $\fnl$ using the well-known {\it scale-dependent bias} effect \cite{dalal/etal:2008, slosar/etal:2008, 2022JCAP...11..013B}. Other ways to constrain CIP using large-scale structure data include methods based on probing directly the local baryon fraction \cite{2009PhRvD..80f3535G, 2010ApJ...716..907H}, mass- vs.~luminosity-weighted galaxy power spectra \cite{2019MNRAS.485.1248S}, and spatial modulations of baryon acoustic oscillations \cite{2019PhRvD.100f3503H, 2021PhRvD.104f3536H}.

\vspace{1mm}
In this paper, we utilize the BOSS DR12 galaxy power spectrum to constrain CIP using the contribution $\propto \bsigma\sigma(\vx)$ in the galaxy bias expansion. The constraints depend on whether the CIP are correlated or uncorrelated with $\R(\vx)$, and we show results for both cases. Further, the galaxy data is sensitive to the product $A\bsigma$, where $A$ is the CIP amplitude, but $\bsigma$ is an uncertain galaxy bias parameter. We show constraints on $A\bsigma$, as well as on $A$ with assumptions made about $\bsigma$. To the best of our knowledge, these are the first real-data constraints on CIP using the $\propto \bsigma\sigma(\vx)$ contribution. For uncorrelated CIP, we find that the BOSS DR12 data may already improve over the current CMB constraints to set the tightest bounds to date, although this depends on the $\bsigma$ assumptions. We show also joint constraints on $A$ and $\fnl$, which is relevant because of their similar effects \cite{2020JCAP...02..005B, 2019PhRvD.100j3528H, 2020JCAP...07..049B} and inflation models that generate one typically generate also the other.

\vspace{1mm}
The rest of this paper is organized as follows. In Sec.~\ref{sec:analysis} we describe the BOSS DR12 data, the theory model and the sampling strategy we adopt. We discuss our constraints on $A\bsigma$ in Sec.~\ref{sec:sod}. In Sec.~\ref{sec:fixedbias} we present measurements of the $\bsigma$ parameter from simulations, which we subsequently use to constrain directly the amplitude $A$ of CIP. The joint constraints on $A$ and $\fnl$ are shown in Sec.~\ref{sec:joint}. We summarize and conclude in Sec.~\ref{sec:conc}. In App.~\ref{app:triangle} we show a number of additional constraint plots, and in App.~\ref{app:cmb} we collect a number of recent CMB constraints on CIP.

\vspace{1mm}
In all our results we keep the following cosmological parameters fixed as $\Omega_b h^2 = 0.02268$, $\Omega_c h^2 = 0.1218$, $h = 0.6778$, $n_s = 0.9649$, $\A_s = 1.75\times 10^{-9}$ ($k_p = 0.05/{\rm Mpc}$; corresponding to $\sigma_8 = 0.75$). These are the values listed on the left-hand side of Table VII of Ref.~\cite{2022PhRvD.105d3517P}, which were obtained for the same data. We use the {\tt CAMB} \cite{camb} code to evaluate the matter power spectrum and transfer function.

\section{Analysis specifications}
\label{sec:analysis}

In this section we describe the specifications of our CIP constraint analysis, including the observational data and its covariance, the theory model, and the sampling strategy adopted.

\subsection{Data and covariance}
\label{sec:data}

\begin{figure}
\centering

\includegraphics[width=\textwidth]{\pathtofigs 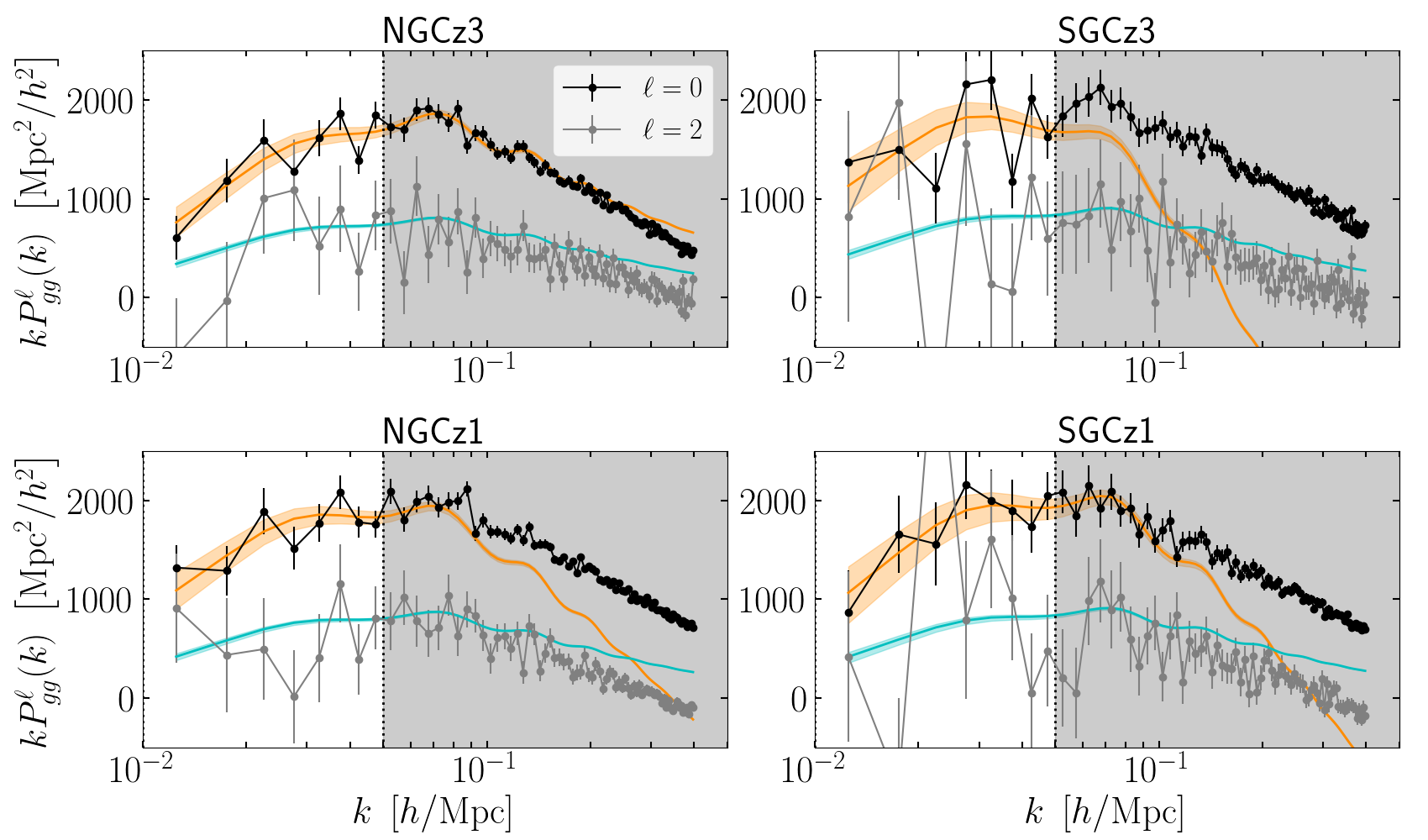}
\caption{Galaxy power spectrum data used in this paper to constrain CIP. The black and grey points with error bars show the monopole ($\ell = 0$) and quadrupole ($\ell = 2$) measurements, respectively; the panels are for the four BOSS DR12 galaxy samples. The grey area marks the data with $k > k_{\rm max} = 0.05 h/{\rm Mpc}$ that we {\it do not} use in the constraints. The orange and cyan lines show the best-fit from our correlated CIP $A\bsigma$ constraints in Sec.~\ref{sec:sod}; the shaded bands around them mark the corresponding $1\sigma$ uncertainty.}
\label{fig:bf}
\end{figure}

We consider the monopole and quadrupole of the BOSS DR12 \cite{2017MNRAS.470.2617A} redshift-space galaxy power spectrum measured with the {\it window-free} method of Ref.~\cite{2021PhRvD.103j3504P}. We utilize the measurements for the galaxy samples at redshifts $z_3 = 0.61$ and $z_1 = 0.38$ in the north (NGC) and south (SGC) galactic caps. We label these samples as \{NGCz3, SGCz3, NGCz1, SGCz1\}, and their volume and total galaxy number are $V = \{2.80, 1.03, 1.46, 0.53\}{\rm Gpc^3}/h^3$ and $N_g = \{435741, 158262, 429182, 174819\}$, respectively. The covariance matrix of the data comes from an ensemble of 2048 {\it MultiDark-Patchy} mock galaxy samples \cite{2016MNRAS.456.4156K, 2016MNRAS.460.1173R}; we apply the correction of Ref.~\cite{2007A&A...464..399H} when we compute the inverse covariance matrix. The data vector is shown by the black (monopole) and grey (quadrupole) data points in Fig.~\ref{fig:bf}.\footnote{We thank Oliver Philcox for these data; \url{https://github.com/oliverphilcox/Spectra-Without-Windows}.}

The theory model described next is based on linear perturbation theory, and to ensure its validity we consider the data measured only up to $k_{\rm max} = 0.05\ h/{\rm Mpc}$. The minimum wavenumber is $k_{\rm min} = 0.01 \ h/{\rm Mpc}$, resulting in 8 $k$ values for each multipole and a total of $N_d = 8 \times 2 \times 4 = 64$ points in our data vector for the four galaxy samples. We do not model observational systematic effects that may affect the large-scale data \cite{2013PASP..125..705P, 2019MNRAS.482..453K, 2021MNRAS.506.3439R}; we return to the importance of this when we discuss our results.

\subsection{Theory model}
\label{sec:model}

Our theory model is based on the following linear galaxy bias expansion (we add $\fnl$ later in Sec.~\ref{sec:joint}),
\bq\label{eq:biasexp1}
\delta_g(\vx, z) = b_1\delta_m(\vx, z) + \bsigma\sigma(\vx) + \eps(\vx),
\eq
where $\delta_g$ is the galaxy number density contrast, $\delta_m$ is the matter density contrast, $\sigma$ is the CIP, and $\eps$ is a stochastic (shot-noise) variable.\footnote{Note that relative perturbations between baryons and CDM are also naturally generated by photon-baryon interactions prior to recombination \cite{tseliakhovich/hirata:2010, barkana/loeb:11, blazek/etal:15, 2016PhRvD..94f3508S, 2016ApJ...830...68A, 2020JCAP...02..005B, 2022MNRAS.511.4333K}. The corresponding bias parameters are $\mathcal{O}(1)$ \cite{2020JCAP...02..005B, 2021JCAP...03..023K}, but these perturbations can be ignored as their amplitude is negligible on the large-scales we use to constrain the primordial CIP.} Note that by the equivalence principle, gravity acts equally on baryons and CDM, and so the large-scale primordial CIP have constant amplitude across cosmic time; that is why $\sigma(\vx)$ does not contain $z$ in its arguments. The bias parameters $b_1$ and $\bsigma$ are the leading-order response of galaxy formation to long-wavelength $\delta_m$ and $\sigma$ perturbations, respectively (see Ref.~\cite{biasreview} for a review on bias). The corresponding power spectrum in redshift-space is given by
\bq\label{eq:Pgg_1}
P_{gg}(k, \mu, z) = \left(b_1 + f\mu^2\right)^2P_{mm}(k,z) + 2\left(b_1 + f\mu^2\right)\bsigma P_{m\sigma}(k,z) + \bsigma^2P_{\sigma\sigma}(k) + \frac{\alpha_P}{\bar{n}_g},
\eq
where $\mu$ is the cosine of the angle of the wavevector $\vk$ with the line-of-sight, $f = {\rm dln}D/{\rm dln}a$ is the structure growth factor, $\bar{n}_g$ is the mean galaxy number density and $\alpha_P$ parametrizes departures of the shot-noise power spectrum from the Poisson expectation $1/\bar{n}_g$. Further, $P_{mm}$ is the linear matter power spectrum, $P_{m\sigma}$ is the matter-CIP cross-power spectrum and $P_{\sigma\sigma}$ is the CIP power spectrum. We follow Ref.~\cite{2020JCAP...07..049B} and parametrize the CIP power spectra in terms of the primordial curvature $\R$ power spectrum $P_{\R\R} = 2\pi^2 \A_s/k^3\left(k/k_{\rm p}\right)^{n_s-1}$ as
\bq\label{eq:spectra_def}
P_{\sigma\sigma}(k) &=& A^2 P_{\R\R}(k), \\
P_{m\sigma}(k,z) &=& \xi \sqrt{P_{mm}(k)P_{\sigma\sigma}(k)} = \frac{3}{5} \M(k,z) \xi A P_{\R\R}(k), 
\eq
where $\delta_m(k,z) = (3/5)\M(k,z)\R(k)$, $\M(k) = (2/3)k^2T_m(k,z)/(\Omega_mH_0^2)$ and $T_m$ is the matter transfer function. The two CIP parameters are $A$ that quantifies the CIP amplitude, and $\xi$ that describes the correlation between $\delta_m$ and $\sigma$. Our theory model can thus be written as
\bq\label{eq:Pgg_2}
P_{gg}(k, \mu, z) = \left[\left(b_1 + f\mu^2\right)^2 + \frac{10 \xi\left(b_1 + f\mu^2\right) A\bsigma}{3\M(k,z)} + \frac{25(A\bsigma)^2}{9\M(k,z)^2}\right]P_{mm}(k,z) + \frac{\alpha_P}{\bar{n}_g}.
\eq
The galaxy power spectrum is sensitive to CIP through the parameter combination $A\bsigma$, i.e., without prior information on $\bsigma$ we can only detect CIP through $A\bsigma$ constraints. In our results below we show constraints on $A\bsigma$, as well as on $A$ by assuming priors on the $\bsigma(b_1)$ relation from simulations. Further, the phenomenology of the constraints depends also on the value of $\xi$. For correlated CIP ($\xi = 1$), the leading-order contribution is $\propto A\bsigma/k^2$, whereas uncorrelated CIP ($\xi = 0$) contribute only on larger scales as $\propto (A\bsigma)^2/k^4$. We show constraints for both these cases.

The predictions for the monopole ($\ell=0$) and quadrupole ($\ell=2$) are obtained as
\bq\label{eq:Pkmulti}
P_{gg}^{\ell}(k,z) = \frac{2\ell+1}{2} \int_{-1}^1 {\rm d}\mu P_{gg}(k,\mu,z) L_{\ell}(\mu),
\eq
where $L_{\ell}(\mu)$ are Legendre polynomials. In this paper we do not account for relativistic effects on large scales, which are expected to be negligible at the constraining power of the BOSS DR12 data \cite{2011JCAP...10..031B, 2010PhRvD..82h3508Y, challinor/lewis:2011, gaugePk, 2016JCAP...05..009R, 2015ApJ...814..145A, 2020MNRAS.499.2598W, 2021JCAP...11..010V, 2022JCAP...01..061C}.

\subsection{Sampling strategy}
\label{sec:results}

We assume the following Gaussian likelihood function (see e.g.~Ref.~\cite{2019MNRAS.486..951W} for the impact of this)
\bq\label{eq:like}
-2{\rm ln}\mathcal{L}(\vtheta) = \left(\vD - \vM(\vtheta)\right)^t \hat{\v{C}}^{-1} \left(\vD - \vM(\vtheta)\right),
\eq
where $\vD$ is data vector (cf.~Fig.~\ref{fig:bf}), $\vM$ is the model prediction for a set of parameters $\vtheta$ (cf.~Eqs.~(\ref{eq:Pgg_2}) and (\ref{eq:Pkmulti})) and $\hat{\v{C}}^{-1}$ is the inverse covariance matrix. We use the {\tt EMCEE} {\tt Python} implementation \cite{2013PASP..125..306F} of the affine-invariant Markov Chain Monte Carlo (MCMC) sampler of Ref.~\cite{2010CAMCS...5...65G} to sample the parameter space. We have validated our constraint methodology by applying it to the mean data vector of the 2048 MultiDark-Patchy mocks and finding that the fiducial value of $A\bsigma = 0$ is recovered comfortably within $1\sigma$ (see Ref.~\cite{2022JCAP...11..013B} for the same validation in the context of $\fnl$ constraints).

For each of the four galaxy samples, we always fit for and marginalize over the values of $b_1$ and $\alpha_P$. We show results for cases where $b_1$ varies freely within wide priors, as well as cases with the following Gaussian priors assumed for them
\bq\label{eq:b1priors}
b_1^{\rm NGCz3} &=& 2.288 \pm 0.15 \ \ \ ,\ \ b_1^{\rm SGCz3} = 2.449 \pm 0.145\ , \nonumber\\
b_1^{\rm NGCz1} &=& 2.172 \pm 0.13 \ \ \ ,\ \ b_1^{\rm SGCz1} = 2.209 \pm 0.14\ ;
\eq
these $b_1$ constraints were obtained in Ref.~\cite{2022PhRvD.105d3517P}~(cf.~their Table VII). Our adoption of these priors serves the purpose to recover some of the constraining power on $b_1$ that is lost by our very conservative choice of $k_{\rm max} = 0.05\ h/{\rm Mpc}$. Importantly, the use of these priors is self-consistent as we adopt the same cosmological parameters as in the corresponding results of Ref.~\cite{2022PhRvD.105d3517P}.

For uncorrelated CIP ($\xi = 0$), the galaxy power spectrum is insensitive to the sign of $A\bsigma$ or $A$, and so for these cases we sample only positive values and quote constraints on $|A\bsigma|$ or $|A|$.

\section{Significance of detection analysis: $A\bsigma$ constraints}
\label{sec:sod}

\begin{figure}
\begin{subfigure}
\centering
\includegraphics[width=0.5\textwidth]{\pathtofigs 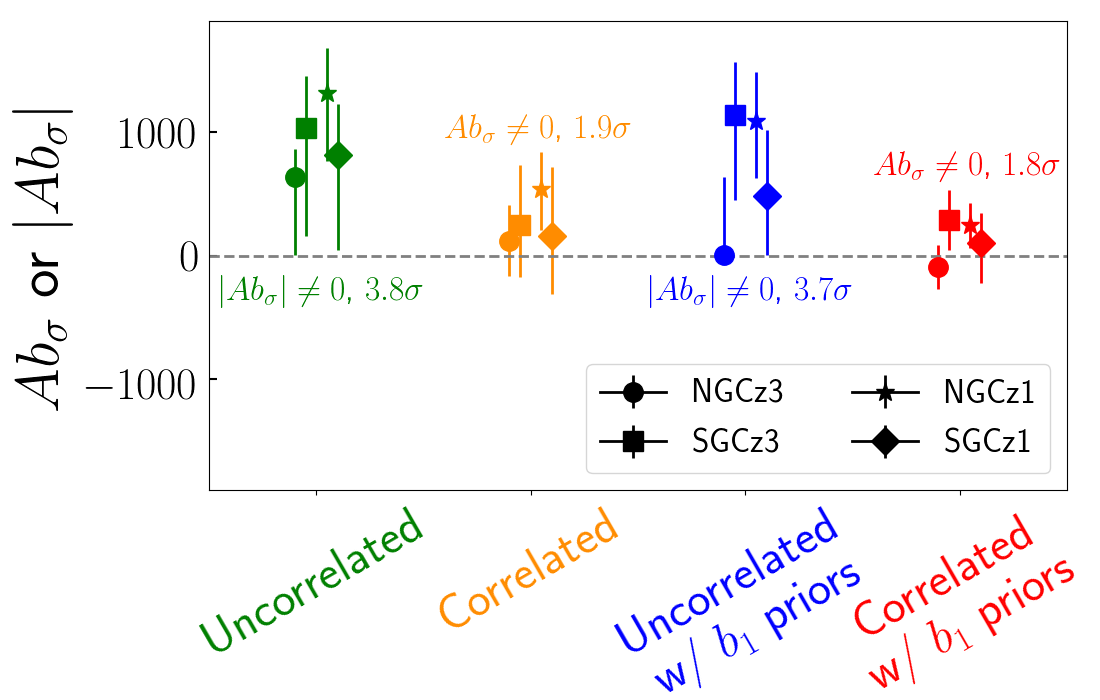}
\end{subfigure}
\begin{subfigure}
\centering
\includegraphics[width=0.5\textwidth]{\pathtofigs 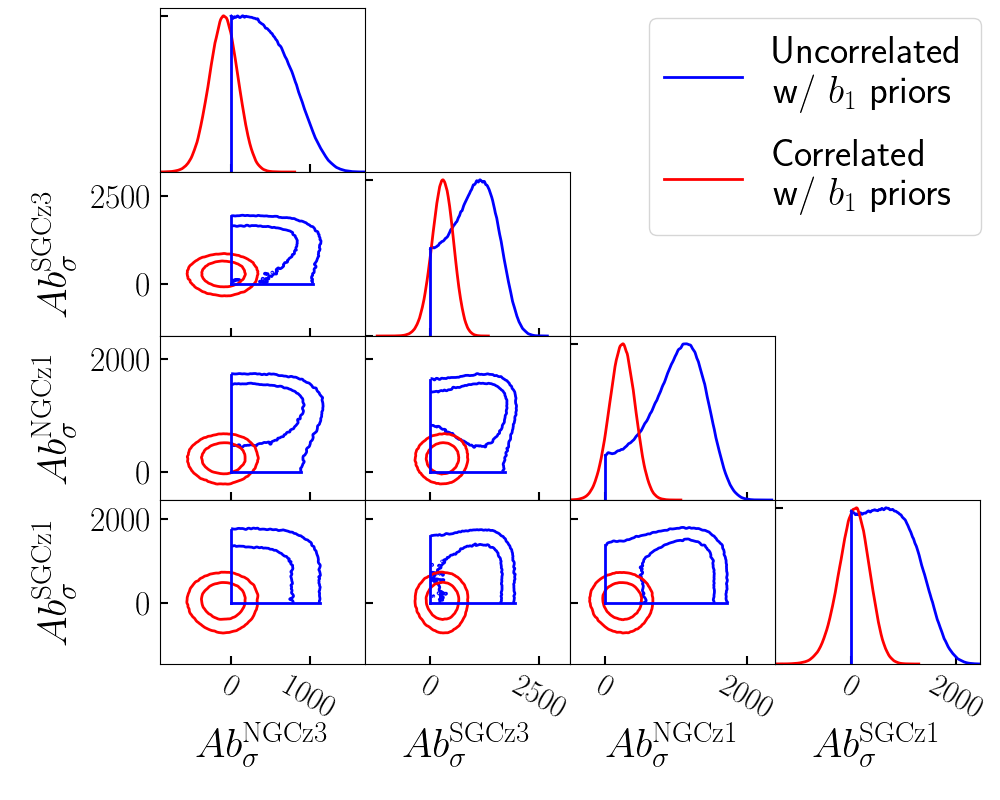}
\end{subfigure}
\caption{Constraints on the parameter combination $A\bsigma$ using the BOSS DR12 galaxy power spectrum. The left panel shows the one-dimensional constraints for correlated ($\xi=1$) and uncorrelated ($\xi=0$) CIP, with and without Gaussian priors on $b_1$; these are also listed in Tab.~\ref{tab:Absigma}. The right panel shows a triangle constraint plot with two-dimensional constraints; note the uncorrelated CIP results are for $|A\bsigma|$. Figure \ref{fig:Absigma_full} in App.~\ref{app:triangle} shows the two-dimensional constraints of the rest of the model parameters.}
\label{fig:Absigma}
\end{figure}

\begin{table*}
\centering
\begin{tabular}{lcccccccccccccc}
\toprule
&  & $A\bsigma^{\rm NGCz3}$ & $A\bsigma^{\rm SGCz3}$ & $A\bsigma^{\rm NGCz1}$ & $A\bsigma^{\rm SGCz1}$ \\
\midrule
\midrule
& Correlated ($\xi = 1$), with $b_1$ priors  & $-93_{-180}^{+180}$ & $287_{-243}^{+243}$ & $250_{-190}^{+172}$ & $98_{-325}^{+248}$  \\
\midrule
& Correlated ($\xi = 1$), no $b_1$ priors  & $117_{-280}^{+296}$ & $244_{-416}^{+485}$ & $535_{-331}^{+306}$ & $157_{-468}^{+562}$  \\
\midrule
& Uncorrelated ($\xi = 0$), with $b_1$ priors  & $ < 637$ & $1138_{-685}^{+430}$ & $1093_{-466}^{+395}$ & $486_{-482}^{+531}$  \\
\midrule
& Uncorrelated ($\xi = 0$), no $b_1$ priors  & $633_{-630}^{+233}$ & $1033_{-878}^{+425}$ & $1317_{-550}^{+362}$ & $814_{-772}^{+413}$  \\
\bottomrule
\end{tabular}
\caption{Constraints on $A\bsigma$ for correlated ($\xi=1$) and uncorrelated ($\xi=0$) CIP using the BOSS DR12 galaxy power spectrum. The quoted errors are $1\sigma$. The constraints for $\xi = 0$ are for the absolute value $|A\bsigma|$.}
\label{tab:Absigma}
\end{table*}

In this section we discuss the constraints on the parameter combination $A\bsigma$, which are independent of assumptions on the bias parameter $\bsigma$. The marginalized $A\bsigma$ constraints are shown in Fig.~\ref{fig:Absigma} and listed in Tab.~\ref{tab:Absigma}; Fig.~\ref{fig:Absigma_full} in App.~\ref{app:triangle} shows a triangle constraint plot for all of the model parameters.

For correlated CIP ($\xi = 1$) we find a significance of detection of $A\bsigma \neq 0$ of $1.8\sigma$ ($1.9\sigma$) for the constraints with (without) Gaussian priors on $b_1$; this is consistent with {\it no detection}. These values are obtained as $\sqrt{{\bf V}_{A\bsigma}^t \cdot {\bf Cov}^{-1}_{A\bsigma} \cdot {\bf V}_{A\bsigma}}$, where ${\bf V}_{A\bsigma}$ is the vector with the mean $A\bsigma$ for the four samples and ${\bf Cov}_{A\bsigma}$ is their covariance estimated from the MCMC. Note that the constraints on $A\bsigma$ are effectively the same as those reported in Ref.~\cite{2022JCAP...11..013B} for the parameter combination $\fnl\bphi$ in local PNG constraints since both impact the galaxy power spectrum in the same way. In fact, any constraint placed on $\fnl\bphi$ that assumes $A = 0$ can be equivalently read as a constraint on $A\bsigma$ for correlated CIP assuming $\fnl = 0$; we return to the degeneracy between local PNG and CIP in Sec.~\ref{sec:joint}.

For uncorrelated CIP ($\xi = 0$) we find a significance of detection of $|A\bsigma| \neq 0$ of $3.7\sigma$ ($3.8\sigma$) for the case with (without) Gaussian priors on $b_1$. This is still consistent with {\it no detection}, but it is substantial enough to analyse with more detail. First, we note that this estimate of the significance of detection assumes the $|A\bsigma|$ parameters are Gaussian distributed, but the right panel of Fig.~\ref{fig:Absigma} shows this is not the case. As a test, instead of sampling $|A\bsigma| > 0$, we sampled $y = (A\bsigma)^2$ with wide linear priors, including allowing it to be negative. The $y$ posteriors were in this case closer to Gaussian, and the significance of detection of $y \neq 0$ lowered to $1.9\sigma$. This shows that the details of the sampling and prior choices have an impact on the reported significance of detection.

Another important aspect of uncorrelated CIP constraints is that, since they are sensitive to larger scales compared to correlated CIP (cf.~Eq.~(\ref{eq:Pgg_2})), they are also more sensitive to potential large-scale data systematics \cite{2013PASP..125..705P, 2019MNRAS.482..453K, 2021MNRAS.506.3439R}. As a simple test, we have checked that removing the lowest $k$ bin is sufficient to bring the significance of detection from $3.7\sigma$ to $\approx 1\sigma$ levels. The exact degree to which systematics in BOSS DR12 may impact our CIP results remains unknown from our analysis, but this is important to quantify in the future.\footnote{This is a message that is relevant also to $\fnl$ constraints obtained with the same galaxy data \cite{2022arXiv220111518D, 2022PhRvD.106d3506C, 2022JCAP...11..013B}.} This is especially so in light of independent Planck data analyses that currently show $\sim 2\sigma$ evidence for uncorrelated CIP \cite{2017JCAP...04..014V, 2020AA...641A..10P}.

\section{Constraints on the CIP amplitude $A$}
\label{sec:fixedbias}

We turn our attention now to the constraints on the CIP amplitude $A$, which require priors on $\bsigma$ to break the $A\bsigma$ degeneracy. We begin by presenting in Sec.~\ref{sec:sims} the simulation measurements of the $\bsigma(b_1)$ relations that we assume. We then discuss the corresponding constraints on $A$ in Sec.~\ref{sec:Aconstraints}.

\subsection{The $\bsigma(b_1)$ relation from separate universe simulations}
\label{sec:sims}

\begin{figure}
\centering
\begin{subfigure}
\centering
\includegraphics[width=0.60\textwidth]{\pathtofigs 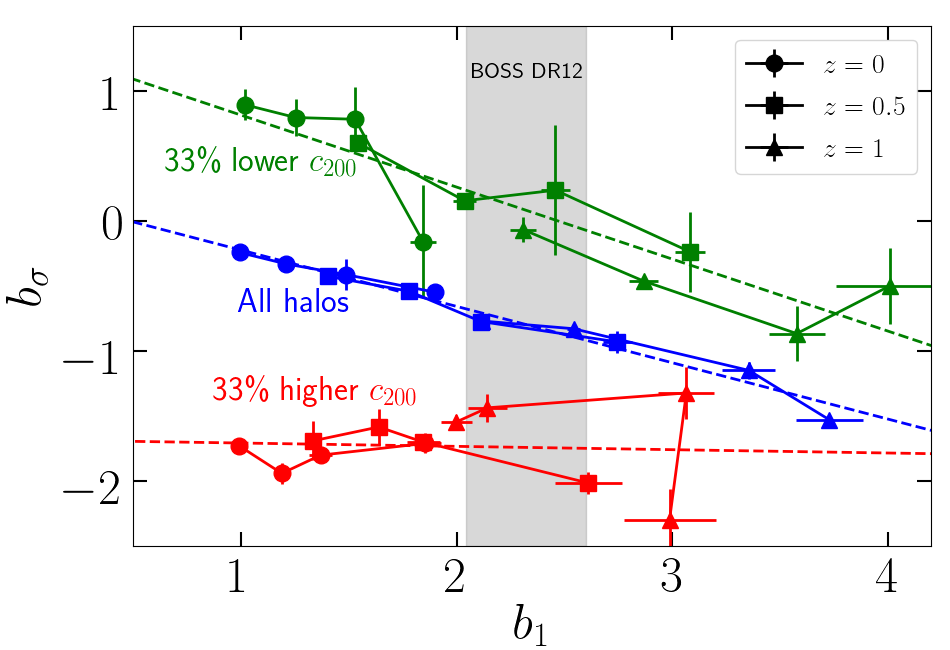}
\end{subfigure}
\begin{subfigure}
\centering
\includegraphics[width=0.49\textwidth]{\pathtofigs 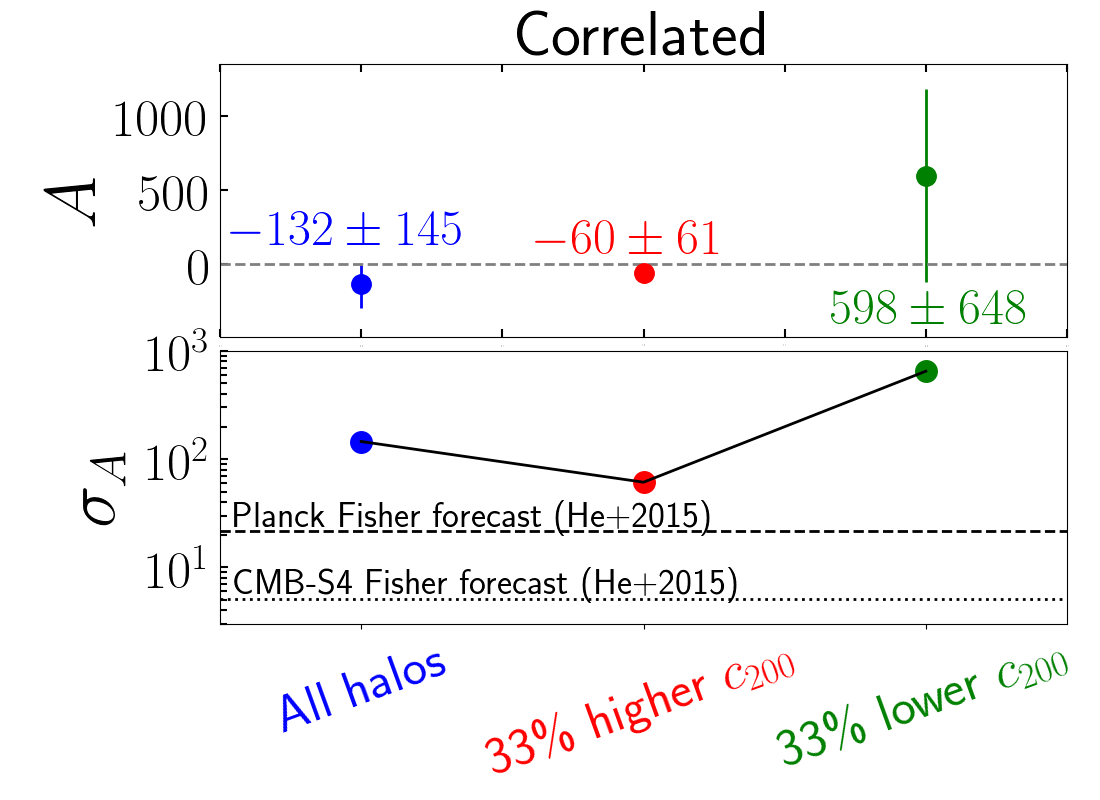}
\end{subfigure}
\begin{subfigure}
\centering
\includegraphics[width=0.49\textwidth]{\pathtofigs 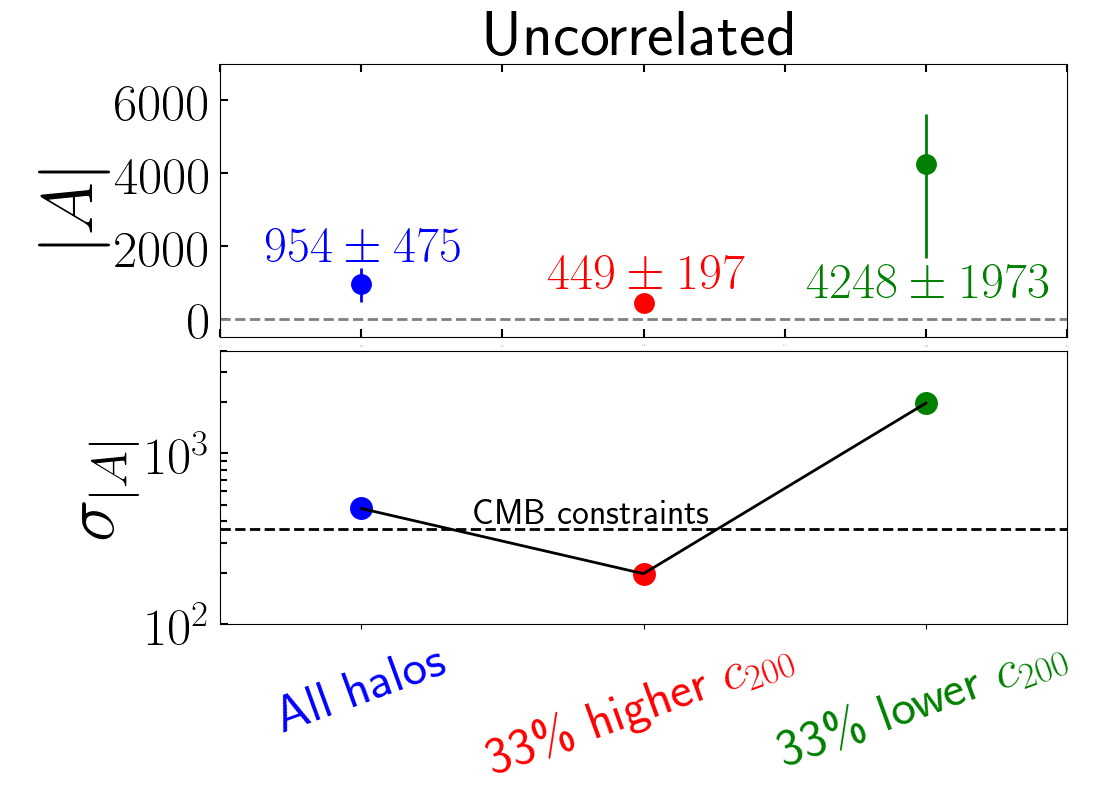}
\end{subfigure}
\caption{Constraints on the CIP amplitude $A$ using the BOSS DR12 galaxy power spectrum for three different assumed $\bsigma(b_1)$ relations. The upper panel shows the $\bsigma(b_1)$ relations, which are for the whole halo population (blue), the $33\%$ most concentrated (red) and $33\%$ least concentrated (green) halos. The symbols with error bars show the simulation measurements at $z=0, 0.5, 1$ (cf.~Sec.~\ref{sec:sims}), and the dashed lines show linear fits to them (cf.~Eq.~(\ref{eq:fits})). The vertical grey band marks the values of $b_1$ expected for the BOSS DR12 galaxies (cf.~Eq.~(\ref{eq:b1priors})). The constraints are shown in the bottom panels for correlated ($\xi = 1$) and uncorrelated ($\xi = 0$) CIP; note that for $\xi = 0$ the constraints are on $|A|$. Marked also are the forecasts for $\sigma_A$ obtained in Ref.~\cite{2015PhRvD..92f3018H} for correlated CIP on the left, and the current CMB constraints for uncorrelated CIP on the right. The result is for the constraints with Gaussian priors on $b_1$; Tab.~\ref{tab:bsigmab1} lists also the constraints without $b_1$ priors. The constraints on the full parameter space are shown in Figs.~\ref{fig:bsigma_full_x1} and \ref{fig:bsigma_full_x0} in App.~\ref{app:triangle}.}
\label{fig:bsigmab1}
\end{figure}

We estimate the bias parameters $b_1$ and $\bsigma$ from a set of gravity-only simulations run with the {\sc Arepo} code \cite{2010MNRAS.401..791S}. The simulation box size is $L_{\rm box} = 560 {\rm Mpc}/h$ with a tracer particle number of $N_p = 1250^3$. The cosmological parameters are the same as the IllustrisTNG galaxy formation simulations \cite{Pillepich:2017jle}: $\Omega_b =0.0486$, $\Omega_m=0.3089$, $\Omega_\Lambda = 0.6911$, $h=0.6774$, $n_s=0.967$, $\sigma_8 = 0.816$. The mass resolution is $m_{\rm p} = 7.7\times 10^{9}\ M_{\odot}/h$. In addition to this fiducial cosmology, we ran also two {\it separate universe simulations} with different values of the baryon $\Omega_b$ and CDM density $\Omega_c$. One is dubbed $\highcip$ and has $\Omega_b^{\highcip} = \Omega_b^{\fidu}\left[1 + \sigma\right]$ and $\Omega_c^{\highcip} = \Omega_c^{\fidu}\left[1 - f_b\sigma\right]$, while the other called $\lowcip$ has $\Omega_b^{\lowcip} = \Omega_b^{\fidu}\left[1 - \sigma\right]$ and $\Omega_c^{\lowcip} = \Omega_c^{\fidu}\left[1 + f_b\sigma\right]$; in these expressions $f_b = \Omega_b^{\fidu}/\Omega_c^{\fidu}$ and we consider $\sigma = 0.05$. Note that all cosmologies have the same total matter density $\Omega_m = \Omega_b + \Omega_c$. 

For all simulations, we identify halos using the FoF algorithm that runs on the fly with the {\sc Arepo} code. The bias parameter $b_1$ is estimated from the fiducial simulation using the large-scale ratio of the halo-matter cross-power spectrum $P_{hm}(k)$ and matter power spectrum $P_{mm}(k)$, 
\bq\label{eq:b1est}
b_1 = \lim_{k \to 0} \frac{P_{hm}(k)}{P_{mm}(k)}.
\eq
We fit this ratio up to $k = 0.15 h/{\rm Mpc}$ with the polynomial $b_1 + Ck^2$, where $C$ is a parameter that absorbs the leading-order deviations from a constant in the mildly nonlinear regime; we have checked our $b_1$ measurements are robust to simply fitting a constant, as well as to the exact maximum $k$ used. As error bars on $b_1$ we consider the error from the least-squares-fit procedure. Concerning $\bsigma$, we estimate it using the separate universe simulations as
\bq\label{eq:bsigmaest}
\bsigma = \frac{\bsigma^{\highcip} + \bsigma^{\lowcip}}{2}, \ \ \ \bsigma^{\highcip} = \frac{1}{+|\sigma|} \left[\frac{n^{\highcip}}{n^{\fidu}} - 1\right], \ \ \ \bsigma^{\lowcip} = \frac{1}{-|\sigma|} \left[\frac{n^{\lowcip}}{n^{\fidu}} - 1\right],
\eq
where $n$ denotes the number density of halos in bins of some halo property (we consider mass and concentration below), and the superscripts label the simulation where the number density is evaluated. These expressions follow from the definition of $\bsigma$ as the response of the halo number density to large-scale CIP, $\bsigma = {\partial {\rm ln} n}/{\partial \sigma}$, which through the peak-background split (PBS) argument \cite{2020JCAP...02..005B, 2021JCAP...03..023K} can be recast as the response to changes to $\Omega_b$ at fixed $\Omega_m$. We take the error bar on $\bsigma$ to be half the difference between $\bsigma^{\highcip}$ and $\bsigma^{\lowcip}$, which should be the same up to numerical noise.

The constraining power on the CIP amplitude $A$ depends on the assumed $\bsigma(b_1)$ relation, which is currently not known for the BOSS DR12 galaxies. Here, we identify a few plausible options for $\bsigma(b_1)$ by considering the impact of halo mass and concentration on these bias parameters. As halo mass definition we take the mass $M_{200}$ inside the radius $R_{200}$ that encloses a mean halo density that is $200$ times the critical cosmic matter density. As halo concentration definition, we take the Navarro-Frenk-White \cite{Navarro:1996} parameter $c_{200}$ measured as proposed in Ref.~\cite{Prada:2011}.

Our measurements of the halo $\bsigma(b_1)$ relation are shown in the upper panel of Fig.~\ref{fig:bsigmab1}. The result in blue is for the whole halo population: the different sets of symbols are for $z=0, 0.5, 1$, and each shows the result in four mass bins in the range $M_{200} \in \left[0.7; 8\right]\times10^{13} M_{\odot}/h$. The green and red symbols show the same, but restricting in each mass bin to the $33\%$ least and $33\%$ most concentrated halos, respectively. The results for all halos agree with those in Refs.~\cite{2020JCAP...02..005B, Voivodic:2020bec, 2021JCAP...03..023K}.\footnote{Inside positive CIP, structure formation is slower because of the larger fraction of baryons that is pressure-coupled to photons until recombination. This explains why $\bsigma < 0$; see Refs.~\cite{2020JCAP...02..005B, Voivodic:2020bec} for more details about the impact of CIP on the formation of halos and galaxies in hydrodynamical simulations.} The strong impact of halo concentration on $\bsigma(b_1)$ is new to this paper. Compared to all halos, the $\bsigma(b_1)$ relation becomes more negative for the halos with higher $c_{200}$, while it is pushed upwards for the lower $c_{200}$ halos, becoming even positive at lower $b_1$.  In this paper we focus solely on the consequences to CIP constraints, and defer studies of the structure formation physics behind these results to future work.

Concerning the connection to the BOSS DR12 samples, Ref.~\cite{2017ApJ...840..104S} used stacked lensing profiles for a subset of these galaxies to constrain their typical host halo concentration to be $\sim 2.5 - 5.8$ (cf.~their Tab.~1). For masses $M_{200} = 2-4\times 10^{13} M_{\odot}/h$ that broadly represent the expected values for BOSS DR12 galaxies \cite{2011ApJ...728..126W, 2013MNRAS.429...98P, 2016MNRAS.455.1553R}, we find that the halos in our lowest and highest concentration tertiles have mean concentrations of $c_{200} \approx 4$ and $c_{200} \approx 5.5$, respectively. Here, we use the fact that our halo samples have mean concentrations that are within the uncertainty of the BOSS galaxies' host halos to justify that the three relations depicted in the upper panel of Fig.~\ref{fig:bsigmab1} are three plausible options for the true $\bsigma(b_1)$ relation of the BOSS DR12 galaxies. Next, we will show constraints on $A$ obtained assuming the following linear fits to the simulation relations (dashed lines in the upper panel of Fig.~\ref{fig:bsigmab1}):
\bq
\label{eq:fits}
{\rm All\ halos} \ &:&\ \bsigma(b_1) = 0.21 - 0.43b_1, \nonumber \\
33\%\ {\rm higher}\ c_{200} \ &:&\ \bsigma(b_1) = -1.68  - 0.03b_1, \nonumber \\
33\%\ {\rm lower}\ c_{200} \ &:&\ \bsigma(b_1) = 1.37  - 0.55b_1.
\eq 

We note that the bias parameter $\bsigma$ has been studied previously in Refs.~\cite{2020JCAP...02..005B, Voivodic:2020bec} using separate universe galaxy formation simulations with the IllustrisTNG model; see also Ref.~\cite{2021JCAP...03..023K} for a work based on two-fluid gravity-only simulations. The physics behind the $\bsigma(b_1)$ relation of galaxies are richer than just the impact of mass and concentration on the relation of their host halos. For example, Refs.~\cite{2020JCAP...02..005B, Voivodic:2020bec} found that CIP modify the galaxy stellar-to-total-mass relation, which has an impact on $\bsigma(b_1)$. As a test, we used the simulation data of Ref.~\cite{2020JCAP...02..005B} to measure the $\bsigma(b_1)$ relation of stellar-mass selected galaxies, which we found to yield $\bsigma$ values compatible with our lower halo concentration relation for $b_1 \sim 2 - 2.5$. We ignore galaxy physics complications like these here, and proceed with the three relations in Eq.~(\ref{eq:fits}) as broadly representative of the impact of different $\bsigma(b_1)$ assumptions on CIP constraints. In the future, it would be interesting to go beyond our simple treatment onto more sophisticated and sample-specific priors on the $\bsigma(b_1)$ relation.

\subsection{Constraints on $A$ for different $\bsigma(b_1)$ relations}
\label{sec:Aconstraints}

\begin{table*}
\centering
\begin{tabular}{lcccccccccccccc}
\toprule
&  & All halos  & $33\%$ higher $c_{200}$ & $33\%$ lower $c_{200}$\\
\midrule
\midrule
& Correlated ($\xi = 1$), with $b_1$ priors  & $-132_{-164}^{+127}$ & $-61_{-64}^{+59}$ & $599_{-717}^{+580}$   \\
\midrule
& Correlated ($\xi = 1$), no $b_1$ priors  & $-297_{-367}^{+248}$ & $-162_{-110}^{+102}$ & $1026_{-413}^{+298}$   \\
\midrule
& Uncorrelated ($\xi = 0$), with $b_1$ priors  & $955_{-501}^{+451}$ & $449_{-249}^{+145}$ & $4248_{-2576}^{+1370}$   \\
\midrule
& Uncorrelated ($\xi = 0$), no $b_1$ priors  & $1150_{-619}^{+526}$ & $547_{-233}^{+165}$ & $3017_{-1360}^{+1275}$   \\
\bottomrule
\end{tabular}
\caption{Constraints on the CIP amplitude $A$ for correlated ($\xi=1$) and $|A|$ for uncorrelated ($\xi = 0$) CIP using the BOSS DR12 galaxy power spectrum assuming the $\bsigma(b_1)$ relations shown in the upper panel of Fig.~\ref{fig:bsigmab1}. The quoted errors are $1\sigma$.}
\label{tab:bsigmab1}
\end{table*}

The lower two panels in Fig.~\ref{fig:bsigmab1} show the constraints on the CIP amplitude $A$ assuming the three $\bsigma(b_1)$ relations in the upper panel; the left and right panels are for correlated ($\xi = 1$) and uncorrelated ($\xi = 0$) CIP, respectively. We assume zero uncertainty on the assumed $\bsigma(b_1)$ relations, i.e.~the only uncertainty on $\bsigma$ is that propagated from $b_1$. The bounds on $A$ are also listed in Tab.~\ref{tab:bsigmab1}, and the constraints on the full parameter space are shown in Figs.~\ref{fig:bsigma_full_x1} and \ref{fig:bsigma_full_x0} in App.~\ref{app:triangle}.

As expected, the assumed $\bsigma(b_1)$ relations control directly the strength of the constraints on $A$. The relation for the $33\%$ most concentrated halos has the largest absolute value of $|\bsigma|$, which leads to the tightest constraints. On the other hand, for the typical $b_1 \sim 2-2.5$ range of values of the BOSS galaxies, we have that $|\bsigma| \ll 1$ for the $33\%$ least concentrated halos, which drastically weakens the constraints. Which of these $\bsigma(b_1)$ relations (if any) describes well the true relation of the BOSS DR12 galaxies is currently uncertain. This prevents us from determining the true constraining power of the data, and so we focus primarily on the impact of different $\bsigma(b_1)$ relations. This is analogous to the situation for $\fnl$ constraints, which are affected by uncertain bias relations as well \cite{2022JCAP...11..013B}.

The bottom left panel compares our three bounds on $A$ for correlated CIP with the CMB data forecasts of Ref.~\cite{2015PhRvD..92f3018H}. According to Ref.~\cite{2015PhRvD..92f3018H}, existing data from the Planck satellite could be used to constrain correlated CIP with a precision of $\sigma_{A} \approx 20$ (with a different method, Ref.~\cite{2021JCAP...08..046C} finds that Planck data could achieve $\sigma_A = 100$). Our strongest constraint assuming that BOSS galaxies are representative of the higher concentration halos is a factor of 3 worse, $\sigma_A \approx 60$ (red). If the lower concentration halos are what is representative of BOSS DR12, then the constraints blow up to $\sigma_{A} \approx 650$. The significance of detection of $A \neq 0$ is $\lesssim 1\sigma$ for all three $\bsigma(b_1)$ relations. This is smaller than the $1.8\sigma$ values we found in our $A\bsigma$ constraints for correlated CIP in Sec.~\ref{sec:sod}. This is not surprising because in $A\bsigma$ constraints each galaxy sample contributes with independent information, whereas in $A$ constraints there is a correlation by the same assumed $\bsigma(b_1)$ relation.\footnote{Concretely, the stronger significance of detection of $A\bsigma$ in Fig.~\ref{fig:Absigma} is largely driven by the NGCz1 sample. When constraining $A\bsigma$, this sample is {\it free} to manifest its {\it preference} for $A\bsigma$ independently of the other samples. However, when constraining $A$, this sample's preference for larger $A$ is balanced by the other samples' preference for lower values.}

The lower right panel of Fig.~\ref{fig:bsigmab1} compares our uncorrelated CIP constraints with the current CMB bounds $|A| \lesssim 360$ (see App.~\ref{app:cmb}). Assuming the BOSS galaxies are representative of all halos (blue) we find $\sigma_{|A|} = 475$, which is a slightly worse precision. Interestingly, should these galaxies represent instead our higher concentration halos, then their power spectrum is able to improve already over the CMB precision with $\sigma_{|A|} = 197$. As expected by the smaller values of $|\bsigma|$, the constraints assuming the $33\%$ least concentrated halos are the weakest with $\sigma_{|A|} = 1973$.

The significance of detection of $|A| \neq 0$ for uncorrelated CIP is $\approx 1.8\sigma$, which is larger than for the correlated case. This is as in our $A\bsigma$ constraints in Sec.~\ref{sec:sod}, where we alerted that prior choices and foreground systematics can have an impact on the significance of detection of uncorrelated CIP. We have checked that removing the lowest $k$ bin brings the significance of detection down to $\approx 1.2\sigma$. Further, this significance of detection of $|A| \neq 0$ for uncorrelated CIP is significantly smaller than that of $\approx 3.7\sigma$ reported in the previous section for $|A\bsigma| \neq 0$. This has to do again with the fact that in $|A|$ constraints the four galaxy samples contribute with correlated information.

\section{Joint constraints on $A$ and local $\fnl$}
\label{sec:joint}

Primordial CIP and local PNG can both be generated in multi-field inflation and contribute in similar ways to the galaxy power spectrum. Accounting for $\fnl$ in the bias expansion of Eq.~(\ref{eq:biasexp1}) \cite{slosar/etal:2008, mcdonald:2008, giannantonio/porciani:2010, 2011JCAP...04..006B, assassi/baumann/schmidt},
\bq\label{eq:biasexpfNL}
\delta_g(\vx, z) = b_1\delta_m(\vx, z) + \bsigma\sigma(\vx) + \bphi\fnl\phi(\vx) + \eps(\vx),
\eq
the redshift-space galaxy power spectrum can be written as
\bq\label{eq:Pgg_fNL}
P_{gg}(k, \mu, z) &=& \Bigg[\left(b_1 + f\mu^2\right)^2 + \frac{2\left(b_1 + f\mu^2\right) \big((5/3) \xi A\bsigma + \fnl\bphi\big)}{\M(k,z)} + \frac{\xi(10/3)A\bsigma\fnl\bphi}{\M(k,z)^2} \nonumber \\
&& + \frac{(5A\bsigma/3)^2 + (\fnl\bphi)^2}{\M(k,z)^2}\Bigg]P_{mm}(k,z) + \frac{\alpha_P}{\bar{n}_g},
\eq
where $\phi = (3/5)\R$ and $\bphi$ is the galaxy bias parameter associated with large-scale primordial $\phi$ perturbations. This equation shows that if $\xi = 1$, then the data is sensitive to the parameter combination $(5/3)A\bsigma + \fnl\bphi$, i.e., the effects of local PNG and correlated CIP are perfectly degenerate. Reference \cite{2020JCAP...07..049B} proposed a way to break this degeneracy by utilizing two galaxy samples with different bias parameters. This allows to constrain $(5/3)A\bsigma^{i} + \fnl\bphi^{i}$ and $(5/3)A\bsigma^{j} + \fnl\bphi^{j}$, where $i$ and $j$ label two galaxy samples, which allows to solve for $A$ and $\fnl$ (see also Ref.~\cite{2019PhRvD.100j3528H} for another way to break the degeneracy based on tomography). Note this requires knowledge of the bias parameters $\bphi$ and $\bsigma$ and does not work for constraints on $\fnl\bphi$ and $A\bsigma$. Here, we break the degeneracy by simply assuming a Gaussian prior with the Planck constraint $\fnl = -0.9 \pm 5.1\ (1\sigma)$ \cite{2020A&A...641A...9P}. For uncorrelated CIP, we do not assume this prior as for $\xi = 0$ the different scale-dependence of the $\fnl$ ($\sim 1/k^2$) and $A$ ($\sim 1/k^4$) terms naturally allows for simultaneous constraints.

We further need to assume a $\bphi(b_1)$ relation in order to break the $\fnl\bphi$ degeneracy and constrain $\fnl$. Here, we utilize the measurements made in Ref.~\cite{2023JCAP...01..023L} using the same approach as in here using separate universe simulations and for the same tertiles of the halo concentration-mass relation. Concretely, we follow Ref.~\cite{2023JCAP...01..023L} and parameterize $\bphi(b_1) = 2\delta_c(b_1 - p)$ with $p = 1, -0.5$ and $2.5$ for all halos, the $33\%$ most concentrated and the $33\%$ least concentrated halos, respectively. 

\begin{figure}
\centering
\includegraphics[width=\textwidth]{\pathtofigs 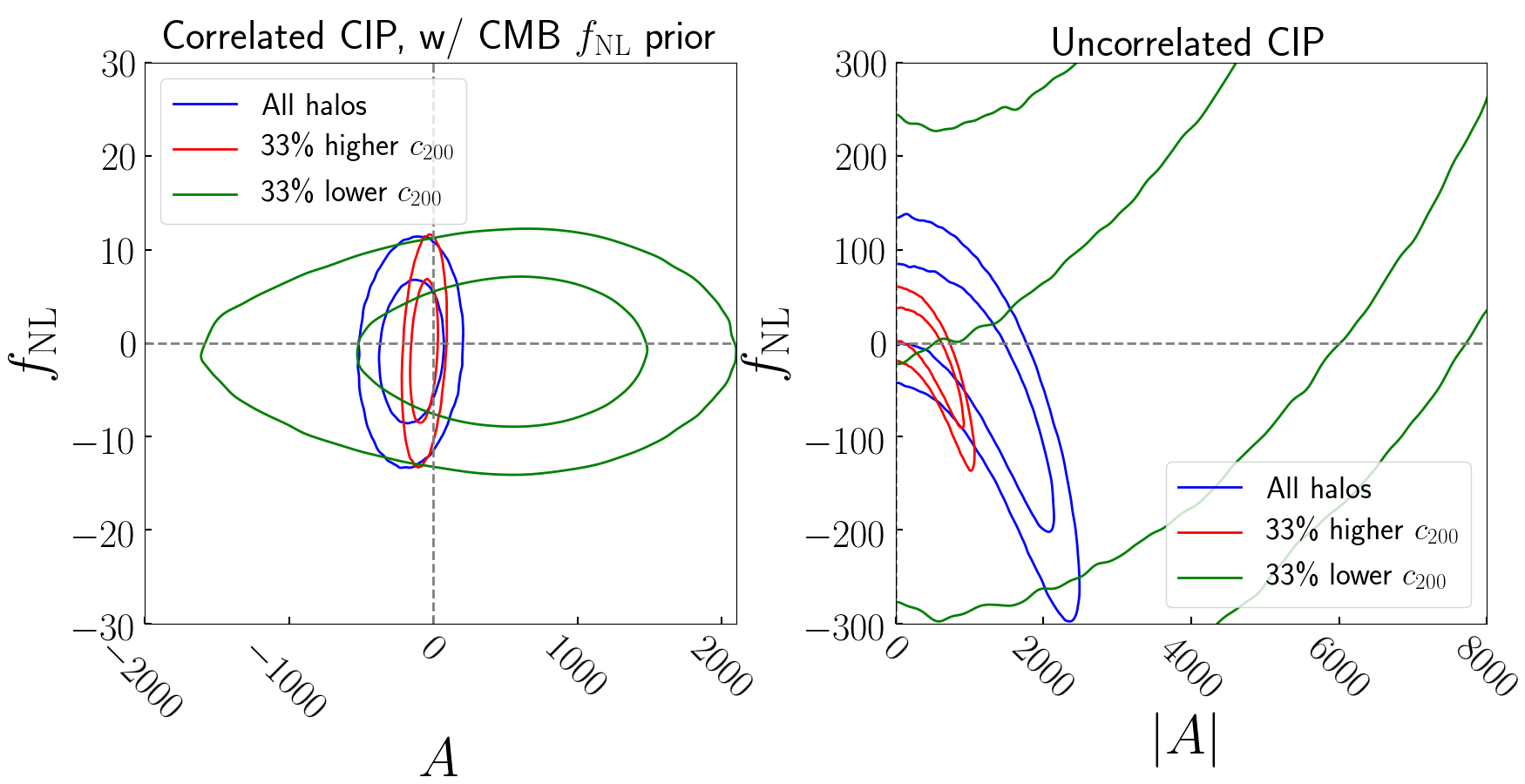}
\caption{Joint constraints on $\fnl$ and $A$ using the BOSS DR12 galaxy power spectrum for different bias parameter relations. For $\bsigma(b_1)$ we consider the same three relations as in Fig.~\ref{fig:bsigmab1}, and for $\bphi(b_1)$ we utilize the relations obtained in Ref.~\cite{2023JCAP...01..023L} for the same halo populations. The left panel is for correlated CIP ($\xi = 1$) with a CMB prior on $\fnl$ to break the perfect degeneracy between $\fnl$ and $A$. The right panel is for uncorrelated CIP ($\xi = 0$), which requires no $\fnl$ prior. The result in both panels is for the case with Gaussian priors on $b_1$. The corresponding one-dimensional constraints are listed in Tab.~\ref{tab:joint}.}
\label{fig:joint}
\end{figure}

The constraints on $\fnl$ and $A$ are shown in Fig.~\ref{fig:joint} and Tab.~\ref{tab:joint}. For correlated CIP, the constraints on $A$ are nearly indistinguishable from those in the previous section assuming $\fnl = 0$. That means the degeneracy between $\fnl$ and $A$ does not degrade the correlated CIP constraints if $\fnl$ is kept within the values allowed by Planck. On the other hand, we do observe a significant degradation of the uncorrelated CIP constraints that do not assume any priors on $\fnl$: compared to the $\fnl = 0$ case, the error bars $\sigma_{|A|}$ increase by $\approx 60\%$. We have also checked (not shown) that if a CMB prior on $\fnl$ is imposed, then there is also no degradation of the uncorrelated CIP constraints.

\begin{table*}
\centering
\begin{tabular}{lcccccccccccccc}
\toprule
&  &  & Correlated ($\xi = 1$), with CMB $\fnl$ prior & \\
\rule{0pt}{3ex}  
&  & All halos  & $33\%$ higher $c_{200}$ & $33\%$ lower $c_{200}$\\
\midrule
& $A$  & $-133_{-162}^{+133}$ & $-65_{-62}^{+64}$ & $607_{-694}^{+584}$   \\
& $\fnl$  & {\it Planck prior} & {\it Planck prior} & {\it Planck prior}   \\
\midrule
\midrule
\rule{0pt}{3ex}  
&  &  & Uncorrelated ($\xi = 0$), without CMB $\fnl$ prior  & \\
\rule{0pt}{3ex}  
&  & All halos  & $33\%$ higher $c_{200}$ & $33\%$ lower $c_{200}$\\
\midrule
& $|A|$  & $1479_{-947}^{+519}$ & $580_{-371}^{+266}$ & $6062_{-3550}^{+2775}$   \\
& $\fnl$  & $0_{-144}^{+64}$ & $2_{-51}^{+27}$ & $21_{-208}^{+426}$   \\
\bottomrule
\end{tabular}
\caption{Constraints on $\fnl$ and A for correlated ($\xi = 1$, top) and uncorrelated ($\xi = 0$, bottom) CIP. For $\xi = 1$, the $\fnl$ constraints are dominated by the CMB prior $\fnl = -0.9 \pm 5.1$, which we assume to break the degeneracy between $\fnl$ and $A$. The result is for Gaussian priors on $b_1$. The quoted errors are $1\sigma$.}
\label{tab:joint}
\end{table*}

\section{Summary \& Conclusions}
\label{sec:conc}

Primordial CIP are a mixture of isocurvature modes produced during inflation characterized by a compensation of baryon and CDM perturbations that leads to no total matter perturbations (cf.~Eq.~(\ref{eq:cipdef})). Unlike other forms of isocurvature which are tightly constrained by the CMB data, the constraints on CIP are remarkably loose: the CIP power spectrum is still allowed to be 5 orders of magnitude larger than the power spectrum of the curvature perturbations $\R(\vx)$ (cf.~App.~\ref{app:cmb}). This motivates investigating alternative ways to improve upon the CMB towards tighter constraints on CIP, which if detected could be used to rule out standard single-field inflation models.

One way to do so is to use the large-scale distribution of galaxies. Concretely, the bias expansion of the galaxy density contrast contains a term $\delta_g(\vx) \supset \bsigma\sigma(\vx)$, where $\bsigma$ is the linear CIP bias parameter that specifies the response of galaxy formation to long-wavelength CIP perturbations $\sigma(\vx)$ (cf.~Eq.~(\ref{eq:biasexp1})). Assuming the CIP and curvature power spectra are related as $P_{\sigma\sigma} = A^2P_{\R\R}$, $P_{\sigma\R} = \xi\sqrt{P_{\sigma\sigma}P_{\R\R}}$, then the galaxy power spectrum acquires contributions $\propto \xi A\bsigma/k^2$ and $\propto (A\bsigma)^2/k^4$, where $A$ is the CIP amplitude and $\xi$ measures the level of correlation between CIP and $\R$ (cf.~Eq.~(\ref{eq:Pgg_2})). This is analogous to the scale-dependent bias effect that can be used to constrain the local PNG parameter $\fnl$ \cite{dalal/etal:2008, slosar/etal:2008, 2022JCAP...11..013B}. In this paper, we relied for the first time on this effect to constrain the amplitude of primordial CIP using the BOSS DR12 galaxy power spectrum.

We assessed the significance of detection of CIP by constraining the parameter combination $A\bsigma$, which can be done independently of assumptions about the uncertain $\bsigma$ parameter. Under a number of plausible assumptions for the relation between $\bsigma$ and the linear density bias parameter $b_1$, we have also constrained the CIP amplitude $A$ directly. In particular, we showed results for three assumed $\bsigma(b_1)$ relations obtained using separate universe simulations for all halos, the $33\%$ most concentrated and $33\%$ least concentrated halos (cf.~Fig.~\ref{fig:bsigmab1}). Our main results can be summarized as follows:

\begin{itemize}

\item The significance of detection of $A\bsigma \neq 0$ is $1.8\sigma$ for correlated ($\xi = 1$) and $3.7\sigma$ for uncorrelated ($\xi = 0$) CIP (cf.~Fig.~\ref{fig:Absigma}). The large significance for uncorrelated CIP should be interpreted carefully in light of the larger sensitivity to potential large-scale data systematics.

\item Concerning the constraints on $A$, the tightest bounds are for the case assuming the $\bsigma(b_1)$ relation of the $33\%$ most concentrated halos: $A = -60 \pm 61\ (1\sigma)$ for correlated and $|A| = 449 \pm 197 \ (1\sigma)$ for uncorrelated CIP (cf.~Fig.~\ref{fig:bsigmab1}). For uncorrelated CIP, this corresponds to an improvement of close to a factor $2$ relative to the current CMB constraints $|A| \lesssim 360\ (1\sigma)$.

\item As expected, however, the constraints on $A$ depend critically on the assumed $\bsigma(b_1)$ relation. For the $\bsigma(b_1)$ of all halos the error bars become a factor of $\approx 2$ worse: $\sigma_A = 145$ for $\xi = 1$ and $\sigma_{|A|} = 475$ for $\xi = 0$ (cf.~Fig.~\ref{fig:bsigmab1}). For the $\bsigma(b_1)$ of the lower concentration halos, the same values are a factor of $\approx 10$ worse than the tighter constraints: $\sigma_A = 648$ and $\sigma_{|A|} = 1973$.

\item Assuming a CMB prior for $\fnl$ leads to no degradation in the $A$ constraints for both correlated and uncorrelated CIP, compared to assuming $\fnl = 0$ (cf.~Fig.~\ref{fig:joint}). If $\fnl$ varies freely there is a degradation of $\approx 60\%$ for uncorrelated CIP. For correlated CIP, the data cannot constrain both $A$ and $\fnl$ without relying on methods involving multiple samples \cite{2019PhRvD.100j3528H, 2020JCAP...07..049B}.

\end{itemize}

Future constraints on CIP using galaxy data will require more work on (i) the impact of large-scale data systematics and (ii) theory priors on the $\bsigma(b_1)$ relation. Concerning the systematics, this is a problem also for $\fnl$ constraints \cite{2013PASP..125..705P, 2019MNRAS.482..453K, 2021MNRAS.506.3439R}, but which is more severe for uncorrelated CIP constraints. The $3.7\sigma$ significance of $|A\bsigma| \neq 0$ for uncorrelated CIP is very interesting from a fundamental physics perspective, but it may also be simply due to systematics in the BOSS DR12 data. This motivates constraining CIP using independent galaxy data, like for example the eBOSS DR16 quasar sample, which has a larger volume and has been subject to more careful large-scale systematic studies \cite{2021MNRAS.506.3439R, 2021arXiv210613725M}. Future CMB survey data are expected to improve upon the current constraints \cite{2017PhRvD..96h3508S}, and may also be used to independently confirm any claimed detection from galaxy data.

Concerning the $\bsigma(b_1)$ relation, without accurate and precise priors for it, it is not possible to constrain $A$ and one must limit to assessing the significance of detection through $A\bsigma$ constraints. This is again analogous to the current situation in $\fnl$ constraints using the scale-dependent bias effect that require priors on the $\bphi(b_1)$ relation \cite{2020JCAP...12..013B, 2020JCAP...12..031B, 2022JCAP...01..033B, 2022JCAP...11..013B, 2022JCAP...04..057B}. A good knowledge of the $\bsigma(b_1)$ relation of real galaxies is useful also to help select galaxy samples with $\bsigma$ values that maximize the significance of detection of CIP; see Ref.~\cite{barreirakrause} for a discussion of this idea in the context of $\fnl$ constraints.

Our results reinforce, with the first real-data analysis, the message from previous forecast studies \cite{2019PhRvD.100j3528H, 2020JCAP...07..049B, 2021PhRvD.103h3519S, 2022arXiv220802829K} that large-scale galaxy data is a very good probe of CIP.  Indeed, if BOSS DR12 galaxies represent higher concentration halos, our results indicate this survey is able to improve already upon current CMB constraints. This is an encouraging message to future surveys like DESI \cite{2013arXiv1308.0847L}, Euclid \cite{2011arXiv1110.3193L} or SphereX \cite{2014arXiv1412.4872D} which will have an even greater constraining power. In the future, it would be interesting also to constrain CIP using the galaxy bispectrum (3-point function).

\acknowledgments
We would like to thank Jos\'{e} Luis Bernal, Giovanni Cabass, Vincent Desjacques, Eiichiro Komatsu, Titouan Lazeyras, Kaloian Lozanov and Fabian Schmidt for very useful comments and conversations. We are also very thankful to Oliver Philcox for making publicly available the BOSS DR12 power spectrum measurements utilized in this paper. The author acknowledges support from the Excellence Cluster ORIGINS which is funded by the Deutsche Forschungsgemeinschaft (DFG, German Research Foundation) under Germany's Excellence Strategy - EXC-2094-390783311. The numerical analysis presented in this work was done on the Cobra supercomputer at the Max Planck Computing and Data Facility (MPCDF) in Garching near Munich.

\appendix 

\section{Additional constraint plots}
\label{app:triangle}

\begin{figure}
\centering
\includegraphics[width=\textwidth]{\pathtofigs 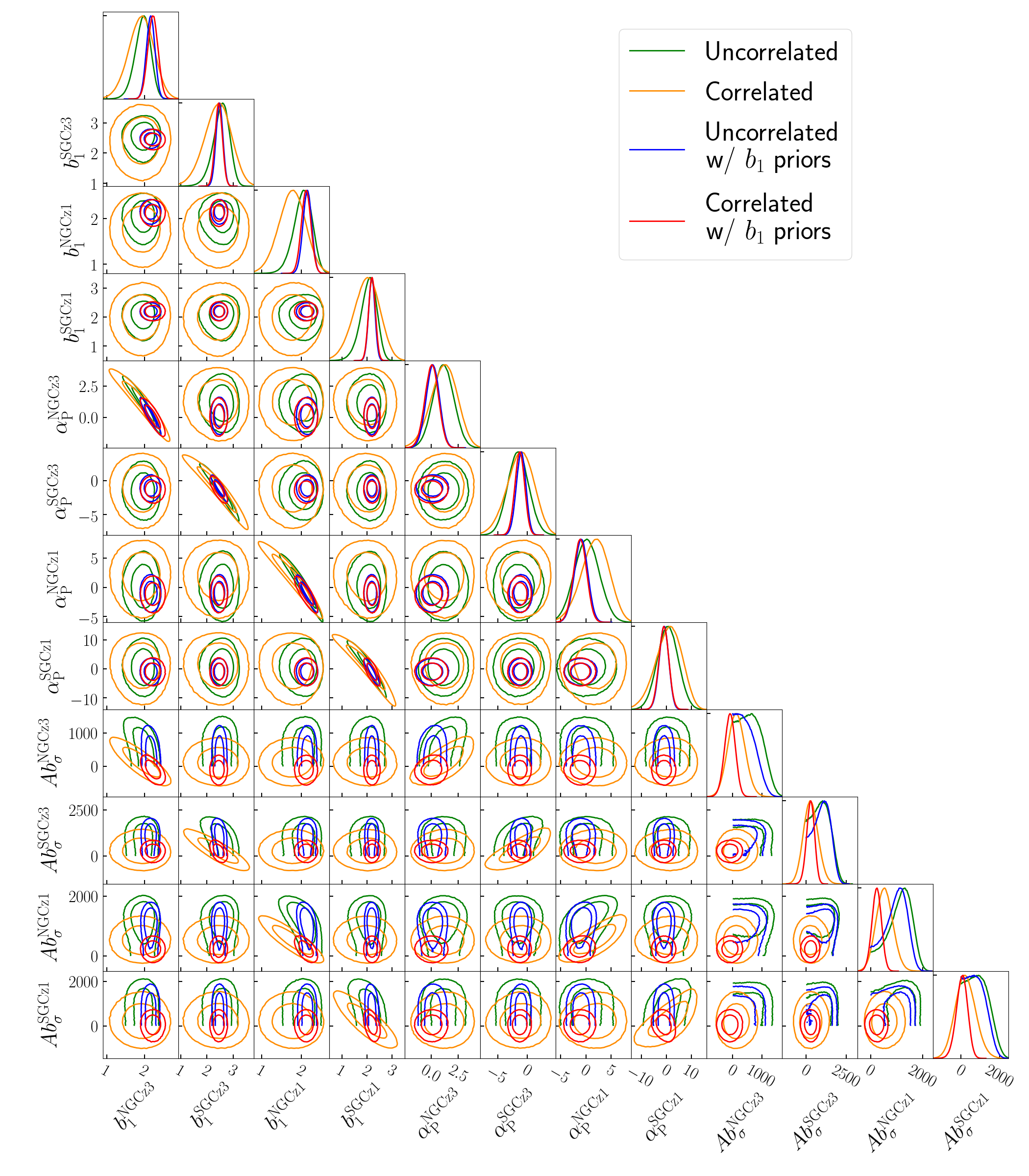}
\caption{Triangle plot with two-dimensional marginalized constraints on the parameters of our $A\bsigma$ constraints. The result is shown for both correlated ($\xi=1$) and uncorrelated ($\xi = 0$) CIP, with and without Gaussian priors on $b_1$. Note that for the uncorrelated CIP cases the constraints are for $|A\bsigma|$.}
\label{fig:Absigma_full}
\end{figure}

\begin{figure}
\centering
\includegraphics[width=\textwidth]{\pathtofigs 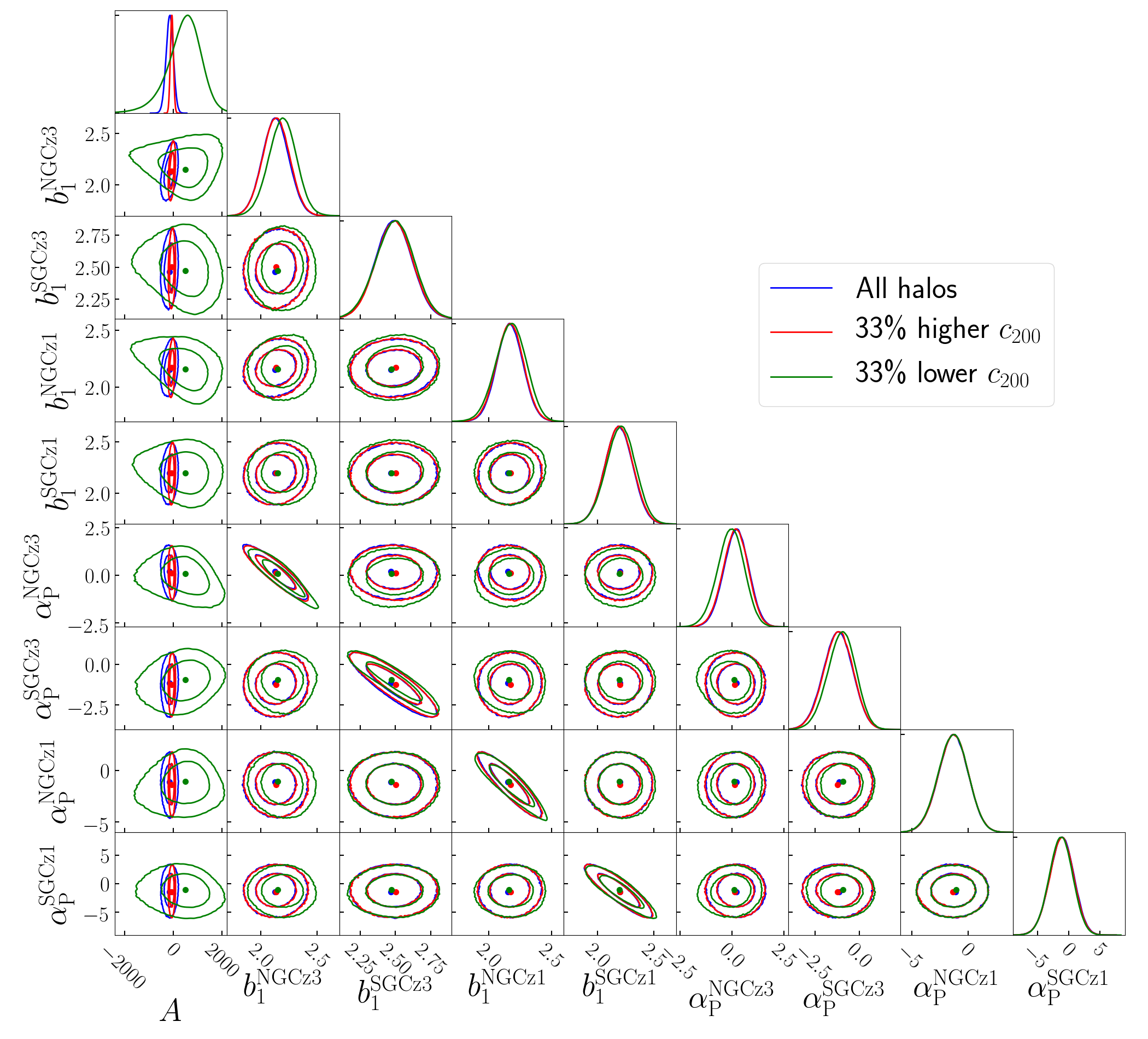}
\caption{Triangle plot with two-dimensional marginalized constraints on the parameters of our constraints on $A$ for different $\bsigma(b_1)$ relations. The result is for correlated ($\xi = 1$) CIP with Gaussian priors on $b_1$.}
\label{fig:bsigma_full_x1}
\end{figure}

\begin{figure}
\centering
\includegraphics[width=\textwidth]{\pathtofigs 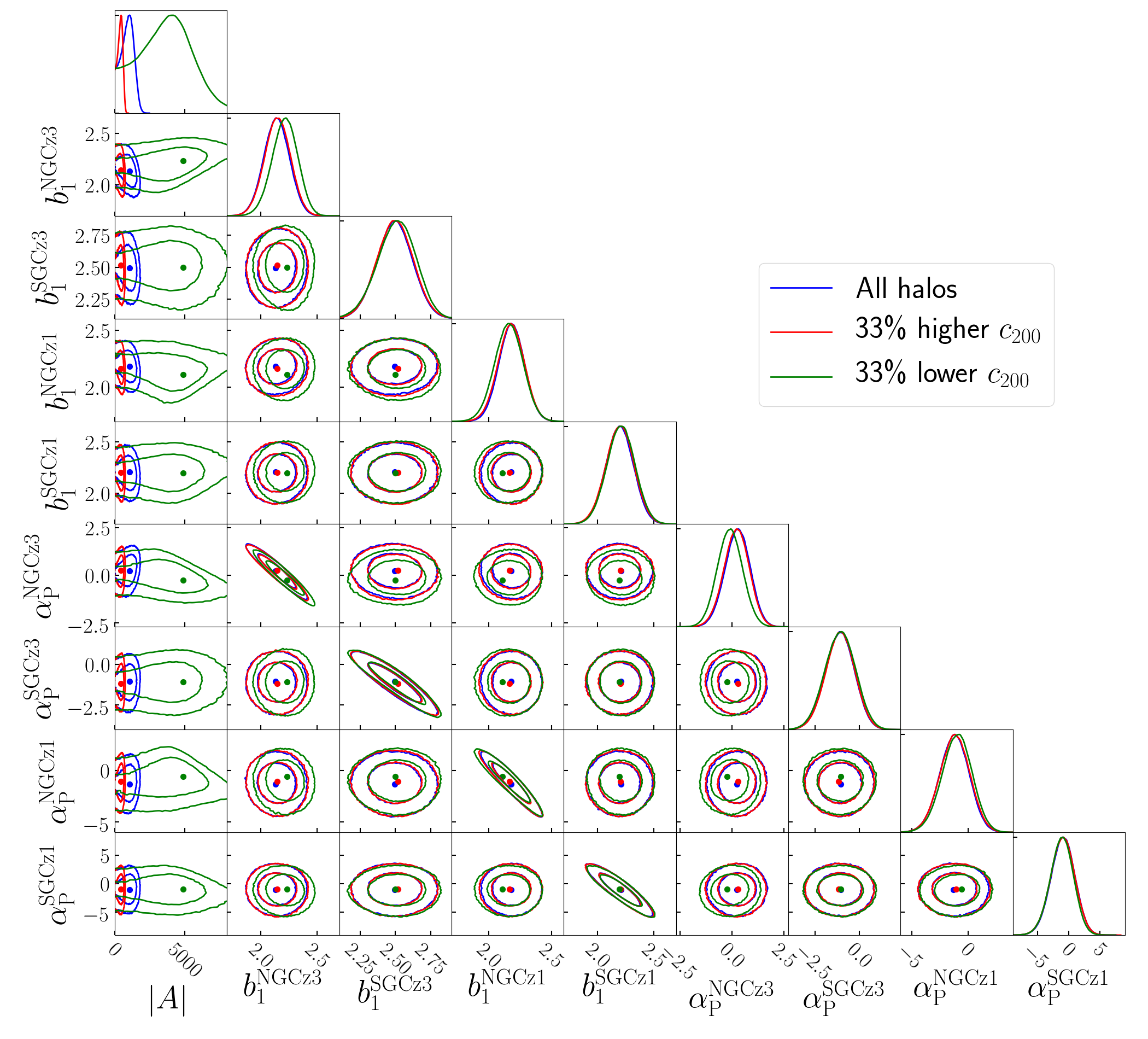}
\caption{Same as Fig.~\ref{fig:bsigma_full_x1}, but for uncorrelated ($\xi = 0$) CIP.}
\label{fig:bsigma_full_x0}
\end{figure}

This appendix displays a few additional plots with two-dimensional parameter constraints. Concretely, 

\begin{itemize}

\item Figure \ref{fig:Absigma_full} shows the constraints on the parameters varied in the analysis of the parameter combination $A\bsigma$ in Sec.~\ref{sec:sod}.

\item Figure \ref{fig:bsigma_full_x1} shows the constraints on the parameters varied in the analysis of the amplitude parameter $A$ for correlated CIP ($\xi = 1$) assuming different $\bsigma(b_1)$ relations.

\item Figure \ref{fig:bsigma_full_x0} shows the same as Fig.~\ref{fig:bsigma_full_x1}, but for uncorrelated CIP ($\xi = 0$).

\end{itemize}

\section{Recent CMB constraints and forecasts}
\label{app:cmb}

Existing CMB constraints on uncorrelated CIP \cite{2016PhRvD..93d3008M, 2017PhRvD..96h3508S, 2017JCAP...04..014V, 2020AA...641A..10P} are quoted in terms of
\bq\label{eq:rmsdef}
\Delta_{\rm rms}^2(R) = \frac{1}{2\pi^2}\int{\rm d}kk^2 \left(\frac{3j_1(kR)}{kR}\right)^2 P_{\sigma\sigma}(k),
\eq
where $j_1(x)$ is the first spherical Bessel function and the CIP power spectrum is assumed scale-invariant, $P_{\sigma\sigma}(k) = A^{\rm SI}/k^3$. Setting $n_s \approx 1$ in Eq.~(\ref{eq:spectra_def}), the relation to our parametrization is $A^{\rm SI} \approx 2\pi^2\mathcal{A}_s A^2 = 3.45 \times 10^{-8} A^2$. To relate the bounds on $\Delta_{\rm rms}^2$ to those on the CIP amplitude, Eq.~(\ref{eq:rmsdef}) can be evaluated from $k_{\rm min} \sim (10\rm Gpc)^{-1}$ assuming $R = R_{\rm CMB} = 148{\rm Mpc}/h$, which gives
\bq
\Delta_{\rm rms}^2(R_{\rm CMB}) = 8.2 \times 10^{-9} A^2.
\eq
The latest constraints are shown in Tab.~\ref{tab:cmb}.\footnote{We mention here also the result from Ref.~\cite{2010ApJ...716..907H} from studying the baryon fraction of galaxy clusters. According to Ref.~\cite{grin/dore/kamionkowski}, this constrains $\Delta_{\rm rms}^2(R = 10{\rm Mpc}/h) < 0.006\ (1\sigma)$, corresponding to $A^2 < 4.64\times 10^5$ ($A < 681$).} They are all consistent with {\it no detection}, despite the hint by Refs.~\cite{2017JCAP...04..014V, 2020AA...641A..10P} for $\Delta_{\rm rms}^2(R_{\rm CMB}) \neq 0$ at the $\sim 2\sigma$ level. The bounds on $A^2$ are $\O(5)$, i.e.~the current CMB constraints allow the uncorrelated CIP power spectrum to be over 5 orders of magnitude larger than the curvature power spectrum $P_{\R\R}$. In the main body of the paper we took half of the $2\sigma$ bound from Ref.~\cite{2017PhRvD..96h3508S}, $|A| \lesssim 360$, as a rough estimate of the typical constraining power of the CMB on uncorrelated CIP (cf.~Fig.~\ref{fig:bsigmab1}); the derivation of more precise bounds needs to take into account the fact that $\Delta_{\rm rms}^2(R_{\rm CMB})$ and $|A|$ have different probability density distributions.

In the future, Ref.~\cite{2017PhRvD..96h3508S} forecasts that CMB data have the potential to probe uncorrelated CIP at the level of $\Delta_{\rm rms}^2(R_{\rm CMB})< 8\times 10^{-5}$, corresponding to $|A| \lesssim 100$.

Regarding correlated CIP, Ref.~\cite{2021JCAP...08..046C} argues that the Planck CMB lensing bispectrum can be used to constrain $A < 100\ (1\sigma)$. Using a different method, Ref.~\cite{2015PhRvD..92f3018H} shows using Fisher forecasts that Planck (CMB-S4) data could be used to probe the CIP amplitude with a precision $\sigma_A = 21$ ($\sigma_A = 5$).

\begin{table*}
\centering
\begin{tabular}{lcccccccccccccc}
\toprule
&  &  $\Delta_{\rm rms}^2(R_{\rm CMB})$  & $A^2$ & $|A|$ \\
\midrule
\midrule
& Mu{\~n}oz+(2016) \cite{2016PhRvD..93d3008M}  & $0.0009 \pm 0.005\ (1\sigma)$ & $(1 \pm 6) \times 10^5\ (1\sigma)$ & $ < 780\ (1\sigma)$   \\
\midrule
& V{\"a}liviita (2017) \cite{2017JCAP...04..014V}  & $0.0069 \pm 0.0030\ (1\sigma)$ & $(8.4 \pm 3.7) \times 10^5\ (1\sigma)$ & $ \in \left[689, 1098\right]\ (1\sigma)$   \\
\midrule
& Smith+ (2017) \cite{2017PhRvD..96h3508S}  & $< 0.0043\ (2\sigma)$ & $ < 5.2 \times 10^5\ (2\sigma)$ & $ < 723\ (2\sigma)$  \\
\midrule
& Planck coll.~(2020) \cite{2020AA...641A..10P}  & $0.0037 \pm 0.0019\ (1\sigma)$ & $(4.5 \pm 2.3) \times 10^5\ (1\sigma)$ & $ \in \left[468, 826\right]\ (1\sigma)$   \\
\bottomrule
\end{tabular}
\caption{Recent CMB constraints on uncorrelated CIP. The constraints are formally placed on the root-mean-square amplitude $\Delta_{\rm rms}^2$; the bounds on $A^2$ and $|A|$ are only approximate (cf.~App.~\ref{app:cmb}).}
\label{tab:cmb}
\end{table*}

\bibliographystyle{utphys}
\bibliography{REFS}

\end{document}